# High-Performance and Low-Power Sub-5 nm Field-Effect Transistors Based on 7-9-7-AGNR

Hang Guo[1], Xian Zhang[2,*], Shuai Chen[3], Li Huang[4,†], Yan Dong[4], and Zhi-Xin Guo[1,‡]

[1.] *State Key Laboratory for Mechanical Behavior of Materials, Center for Spintronics and Quantum System, School of Materials Science and Engineering, Xi'an Jiaotong University, Xi'an, Shaanxi, 710049, China.*
[2.] *Shaanxi Key Laboratory of Surface Engineering and Remanufacturing College of Mechanical and Materials Engineering Xi'an University, Xi'an 710065, China.*
[3.] *Jinan Aviation Repair Factory, No. 600, Zhangjiazhuang, Jinan, 250000, China.*
[4] *State Key Laboratory of Oral & Maxillofacial Reconstruction and Regeneration; National Clinical Research Center for Oral Diseases; Shaanxi International Joint Research Center for Oral Diseases; Department of General Dentistry and Emergency, School of Stomatology, Fourth Military Medical University, Xi'an 710032, China*

\* zhangxian1@outlook.com
†huanglifmmu@qq.com
‡ zxguo08@xjtu.edu.cn

**Abstract**

Recently, an extremely-air-stable one-dimensional 7-9-7-AGNR was successfully fabricated. To further reveal its potential application in sub-5-nm field-effect transistors (FETs), there is an urgent need to develop integrated circuits. Here, we report first-principles quantum-transport simulations on the performance limits of n- and p-type sub-5-nm one-dimensional 7-9-7-AGNR FET. We find that the on-state current ($I_{on}$) in 7-9-7-AGNR FET can be effectively manipulated by the length of the gate and underlap. Particularly, the optimized $I_{on}$ in n-type (p-type) device can reach up to 2423 (4277) and 1988 (920) µA/µm for high-performance and low-power applications, respectively. The large $I_{on}$ values are in the first class among the LD FETs, which can well satisfy the ITRS requirements. We also find that the 7-9-7-AGNR FET can have ultralow subthreshold swing below 60mV/dev, ultrashort delay time (<0.01 ps), and very small power-delay product (<0.01 fJ/µm). Our results show that the 7-9-7-AGNR based FETs have great potential applications in the high-speed and low-power consumption chips.

# I. INTRODUCTION

The scaling of transistors has improved their performance and increased the functionality of the circuit from one generation to the next in the past half-century [1]. While, this trend is ending as the gate length approaches sub-5-nm regimes, because such a short channel suffers from unavoidable high heat dissipation and short-channel effects [2]. To extend the gate length down to the sub-5-nm regime, low-dimensional semiconductors are extensively explored owing to their excellent gate electrostatics in atomically thin crystals and improved carrier transport in the dangling-bond-free smooth surface [3–6]. The low-dimensional semiconductors are expected to hold strong advantages as the candidate channel materials for ultrascaled FETs. By far, numerous efforts have been devoted to exploring two-dimensional (2D) channel materials, such as phosphorene, silicene, tellurene, InSe, $Bi_2O_2Se$, $WSe_2$, GeSe, $ReS_2$, BiN, AsP, $MoSi_2N_4$ for sub-5-nm FETs [7–19]. Nevertheless, most of these materials suffer from the either air stability and/or device performance.

It had been theoretically predicted that the on-state current ($I_{on}$) of monolayer phosphorene (BP) and InSe can satisfy the high-performance (HP) and low-power (LP) requirements of the International Technology Roadmap for Semiconductors (ITRS) in the sub-5-nm region [18,19], and the carrier mobility can reach up to 1000 $cm^2\,V^{-1}\,s^{-1}$ in experimentally fabricated FETs [20–23]. However, their instability in air is an issue, which greatly limits the device fabrications and applications [24,25]. Recently, we found that monolayer $MoSi_2N_4$ is an excellent channel material with both outstanding air stability and device performance [17]. However, its low-cost and large-scale preparation technology suitable for the industrial production is still to be explored. Especially, despite the 2D FETs have been widely studied, the progress on one-dimensional (1D) FETs is far from sufficient. Hence, searching 1D semiconductors with high stability, good FET device performance, as well as simple preparation process is still in great need.

Years ago, graphene nanoribbons attracted great attention for the application as FET channel material due to its very high carrier mobility [5]. Nevertheless, the large-scale preparation of graphene nanoribbons with uniform band gap, which is required for the application in FETs [26-29], is still a challenge. Recently, the on-surface reaction method that can realize high-quality target nanostructures with a minimum of possible side products has been successfully applied to prepare graphene nanoribbons. Massive 7-9-7 armchair graphene nanoribbons (AGNRs) with uniform band gap of 1.8 eV have been synthesized, which are expected to be very stable in air condition with the temperature below 430 ºC [30]. These features make 7-9-7 AGNR has great potential applications as channel material in the FET devices. While, how and whether the its device performance is available remains to be explored.

In this work, we theoretically investigate the performance of n- and p-type sub-5-nm double-gated (DG) one-dimensional 7-9-7-AGNR FET by using *ab initio* quantum-transport calculations. The main critical properties of sub-5-nm FET devices, including the on-state current ($I_{on}$), subthreshold swing (SS), delay time ($\tau$), and power-delay product (PDP), are evaluated. It is found that the optimized $I_{on}$ in both n-type and p-type devices can well satisfy the ITRS requirements. It is also found that the 7-9-7-AGNR based FET holds an ultralow SS, ultrashort $\tau$, and ultrasmall PDP. These results show that the 7-9-7-AGNR is an ideal channel material for the 1D FETs.

## II. METHOD

Geometric optimization of 7-9-7-AGNR is calculated by density-functional theory (DFT) with the projector-augmented-wave (PAW) method, which is implemented in the Vienna ab initio simulation package (VASP) [31-33]. The exchange-correlation function is described based on the generalized gradient approximation (GGA) in the form of Perdew-Burke-Ernzerhof (PBE) parameterization [34]. The convergence standards of the atomic energy and positions are less than $1 \times 10^{-5}$ eV per atom and $1 \times 10^{-2}$ eV Å$^{-1}$, respectively. The cutoff energy of the wave function is set to 500 eV. The Brillouin zone is sampled by a $1 \times 1 \times 3$ Monkhorst-Pack k-point mesh for geometrical optimization [35].

The electronic structure and transport properties are simulated based on the DFT method combined with the nonequilibrium Green's function (NEGF) formalism, using the Atomistix ToolKit (ATK) 2019 package [36,37]. In the NEGF calculations, the drain current at a given bias voltage ($V_{bis}$) and gate voltage ($V_g$) is calculated through the Landauer-Büttiker formula [38]

$$I = \frac{2e}{h}\int_{-\infty}^{+\infty} dE\, T(E, V_{bis})[f_L(E - \mu_L) - f_R(E - \mu_R)] \quad (1)$$

where $T(E, V_{bis}, V_g)$ is the transmission coefficient; $f_L$ and $f_R$ stand for the Fermi-Dirac distribution functions for the source and drain, respectively. $\mu_L$ and $\mu_R$ are the electrochemical potentials of the source and drain, respectively. In the calculations, the *tier-3* basis set is adopted with Hartwigsen-Goedecker-Hutter pseudopotentials and GGA in the form of the PBE function is utilized to represent the exchange and correlation interactions. The real-space mesh cutoff is chosen to be 75 Hartree, and the k-point meshes [39] are set as Monkhorst-Pack 1 × 1 × 150 and 1 × 1 × 1 meshes for the electrode region and the central region in the Brillouin zone (BZ). Moreover, the boundary condition along the transverse, vertical, and transport directions are set to be of periodic, Neumann, and Dirichlet type, respectively [40].

## III. RESULT AND DISCUSSION

### A. Channel materials and device configuration

In set of Fig. 1(a) shows the atomic structure of 1D 7-9-7-AGNR, which has an optimized lattice constant of 26.02 Å in the length (z) direction, consistent with previous theoretical results [30]. The energy band structure of 7-9-7-AGNR is also plotted in Fig. 1(a), which presents a direct band gap of 1.8 eV that is significantly larger than conventional graphene nanoribbons (GNR) of the same width [2]. This result shows that a suitable band gap has been opened up in the 7-9-7-AGNR through the edge engineering. As shown in Fig. 1(b), the analysis of transmission-eigenstate shows that the marginal benzene ring plays an important role in regulating the transport channel. That is, the projection eigen states nearby Fermi level are mainly concentrated around a "zigzag" transport channel formed between two nearest-neighboring marginal benzene rings, which is very different from that in the conventional GNRs.

We additionally evaluate the effective mass that directly are related to the carrier mobility in 7-9-7-AGNR. The electron effective mass ($m_e^*$) around CBM and the hole effective mass ($m_h^*$) around VBM can be calculated with $m^* = \hbar^2 [\partial^2 \varepsilon(k)/\partial k^2]^{-1}$, where $\varepsilon(k)$ is the energy dispersion function corresponding to the band structure. It is found that the effective mass of electron (hole) is 0.151 $m_e$ (0.354 $m_e$). Note that a smaller effective mass ensures a larger on-state current in the device [17].

The performance of 7-9-7-AGNR based FET is further investigated by using the ATK package. In the simulation, a double-gate FET structure is adopted, and degenerately doped 7-9-7-AGNR is selected as two-probe electrodes [Fig. 2(a)]. Compared with the single-gate structure, dual gates can obviously increase the competence of gate modulation [41]. The electron-transport direction along 7-9-7-AGNR is defined as the z direction, and the gates are assumed to be ideal rectangles in the device model. The electrode with a length of about 26.02 Å is adopted, and the silicon dioxide with dielectric constant of 3.9 is used as the dielectric layer. As indicated in Fig. 2(a), in the FET channel the gate region is usually shorter than the dielectric region, and the uncovered dielectric region is called the underlap region. Previous studies found that the gate underlap, which is a spacer area between the gate and electrode, plays an essential role in the scalability of the gate length for the FETs [21]. An appropriate length of the gate underlap ($L_{UL}$), could improve the device's performance. Hence, one can define the channel length ($L_{ch}$) as a sum of the gate length ($L_g$) and twice the length of $L_{UL}$, namely, $L_{ch} = L_g + 2L_{UL}$.

In the calculations, the atomic-compensation-charge method is used for the doping in the electrodes. As indicated in Figs. 2(b) and 2(c), the doping concentration is tested with values of $1.0 \times 10^{13}$, $2.0 \times 10^{13}$, $3.0 \times 10^{13}$, and $4.0 \times 10^{13}$ cm$^{-2}$ [42]. It is seen that the I-V curves with different concentrations are very similar. Note that negative differential electronics (NDR) appears around 0.6 eV and 0.3 eV for the n-type and p-type FETs, respectively. Such NDR can be owing to the unique 1D band structure in the 7-9-7-AGNR, which has an obvious quantized characteristic [43].

In the following studies, we focus on the FET performance with a doping concentration of $1.0 \times 10^{13}$ cm$^{-2}$. According to the ITRS 2013 edition requirements for HP and LP standards of a sub-5-nm device in 2028, we adopt 0.6 V as the supply voltage ($V_{bis}$) and 0.41 nm as the equivalent oxide thickness (EOT) of a dielectric material (silicon dioxide). Various lengths of gate underlap, ranging from 0 to 4 nm, are considered for a comprehensive investigation of the performance of 7-9-7-AGNR FET.

### B. On-state current

The calculated I-V characteristics of sub-5-nm gate-length 7-9-7-AGNR FETs with $L_g$ =1 nm, 3 nm, 5 nm and $L_{UL}$= 0-4 nm, respectively, are shown in Fig. 3. It is seen that all of the FETs with either n-type or p-type doping, have a small enough source-drain leakage current ($I_{off}$) that satisfies the off-state requirement of HP standard ($5 \times 10^{-5}$ μA/μm), and most of the leakage current even satisfies the off-state requirement of LP standard ($0.1 \times 10^{-5}$ μA/μm). The only exception appears in the n-type FETs with $L_g$ =1 nm and $L_{UL}$>= 2 nm [Fig. 3(a)], which does not fully satisfy the requirement of LP standard due to the short-channel effect [44]. The good performance of leakage current in the FETs can be attributed to large band gap and simple energy-band state, as well as the atomic layer thickness of 7-9-7-AGNR that has a strong gate controllability.

We then explore the on-state current ($I_{on}$) of 7-9-7-AGNR FET, which is a key parameter for evaluating the transition speed of a logic device. A high $I_{on}$ is advantageous for efficient applications, such as high-performance servers with a high switching velocity. The HP and LP currents can be calculated by applying a bias of $V_g(\text{on-HP}) = V_g(\text{off-HP}) \pm V_{bis}$ and $V_g(\text{on-LP}) = V_g(\text{off-LP}) \pm V_{bis}$, respectively. As shown in Fig. 3, $V_g(\text{off-HP})$ and $V_g(\text{off-LP})$ are the voltage values where $I_{off}$ reaches $0.1 \times 10^{-5}$ µA/µm and $5 \times 10^{-5}$ µA/µm, respectively, according to the ITRS requirements.

We further summarize the values of $I_{on}$ in Fig. 4. A common feature is that $I_{on}$ can be significantly modulated by the gate length and underlap length in both n-type and p-type FETs. Note that the n-type devices can fully meet the ITRS HP requirement (900 µA/µm). The maximum $I_{on}$ reaches up to 2496, 2195, and 1899 µA/µm with the gate length being 1-, 3- and 5-nm, respectively, which is realized by choosing the proper underlap length [Fig. 4(a)]. As for the p-type FET, the ITRS HP requirement can be met only with the smallest (1 nm) and largest (5 nm) gate length [Fig. 4(b)]. The maximum $I_{on}$ (4277 µA/µm) appears with $L_g$=1 nm and $L_{UL}$=2 nm, which is about two times larger than that of n-type FET. Figs. 4(c) and 4(d) additionally show the variation of $I_{on}$ with gate length and underlap length. In the n-type FET, the LP requirement (300 µA/µm) can be satisfied only with $L_g$>3.5 nm. Whereas, it can be satisfied with the gate length being 1.0-5.0 nm, provided that the proper underlap length is adopted. Different from the HP case, the maximum $I_{on}$ in the LP performance appears in a n-type device, which reaches up to 1988 µA/µm, more than 6 times larger than the ITRS LP requirement. The different behaviors of $I_{on}$ between the n-type and p-type devices are from the combined effects of distinct effective masses between electrons (0.151 $m^*_e$) and holes (0.354 $m^*_e$), and the different electronic density of states between the conduction and valence states, both of which directly correspond to the electronic transport property. Note that the unusually large $I_{on}$ in the n-type LP and p-type HP devices shows that the 7-9-7-AGNR is an ideal candidate for the high-performance FETs.

It should be noted that the increase of underlap length does not always improve the on-state current (as indicated in Fig. 4). This feature can be owing to two competing mechanisms: On the one hand, an increase in the $L_{UL}$ makes the channel barrier longer, which decreases the transmission possibility and suppresses the short-channel effect (positive effect). On the other hand, the gate-controlling capability of the $L_{UL}$ becomes weaker as its length increases, which would degrade the performance of the device (negative effect). These two conflicting effects lead to the nonmonotonic variation of $I_{on}$ with $L_{UL}$. Therefore, $L_{UL}$ should be optimized to obtain the highest $I_{on}$ at a fixed gate length.

To further illustrate the function of underlap, we calculate the $L_{UL}$ dependent local density of states (LDOS) of p-type FET with the gate length fixed at a 1 nm in the LP application (Fig. 5). We define the maximum-electron-barrier height, $\Phi_m$, as the energy barrier for electronic transport from the source to the drain. As shown in Figs.5(a)–5(c), with the same off-state current of $5\times10^{-5}$ μA/μm, $\Phi_m$ is reduced from 0.55 eV at $L_{UL}$ = 0 nm to 0.4 and 0.37 eV at $L_{UL}$ = 2 and 4 nm, respectively. When a gate voltage of 0.6 V is applied, the VBM of 7-9-7-AGNR in the channel region moves upward, leading to the on state of FET [Figs. 4(d)–4(f)]. From Fig. 4(d), $I_{on}$ with $L_{UL}$ = 4 nm is the highest, followed by that of $L_{UL}$ = 2 nm, and $I_{on}$ with $L_{UL}$ = 0 nm is the lowest. This is because the VBM in the channel region becomes higher as $L_{UL}$ increases, which leads to a higher on-state current. This result shows that the variation of $L_{UL}$ can induce various carrier barrier heights ($\Phi_m$), and thus leads to rich gate-control effect.

### C. Gate control

We additionally explore the gate-control ability of the FET in the subthreshold region (denoted as SS), which is another important factor for the device's performance and operating voltage [63]. The definition of SS is

$$SS = \frac{\partial V_g}{\partial log_{10} I_{DS}} \quad (2)$$

where $I_{DS}$ is the drain current. The smaller SS indicates to a better gate-control ability of a FET device. Figure 6 shows the variation of SS with gate length and underlap length. As one can see, SS generally decreases with increasing $L_g$ at a certain $L_{UL}$, showing that the increase of gate length is beneficial for reducing SS. The minimum SS values appear with $L_{UL}$ = 2 nm and 4 nm in the n-type and p-type devices, which are 65 mV/dec and 54 mV/dec, respectively. These values of SS are much smaller than that without underlap (80 mV/dec and 70 mV/dec for the n-type and p-type devices, respectively), showing the strong modulation effect of underlap on the limit of SS. This is because devices with nonzero $L_{UL}$ can help to reach the off-state current with a smaller $V_g$, which leads to a more significant slope in the I-$V_g$ curves. Note that the underlap effect is more noticeable in the relatively-short-gate-length devices. For example, in the p-type FETs with $L_{UL}$ = 4 nm, SS of $L_g$ = 1 nm FET can be reduced by 45% (from 98 to 55 mV/dec), while it can only be reduced by 22% (from 69 to 54 mV/dec) with $L_g$ = 5 nm. The small SS values which are close to or even smaller than the Boltzmann tyranny (60 mV/dec), implies the excellent performance of 7-9-7-AGNR FETs.

### D. Delay time and power consumption

Finally, we reveal the property of switching speed and power consumption in the 7-9-7-AGNR FETs, which are also essential figures of merit for a digital circuit. The switching speed can be characterized directly by the intrinsic delay time ($\tau$), as

$$\tau = \frac{C_g V_{dd}}{I_{on}}. \tag{3}$$

$C_g$ is the total gate capacitance, defined as the sum of the channel capacitance ($C_{ch}$) and the gate fringing capacitance ($C_f$) per width. $C_f$ is speculated to be two times of the intrinsic channel capacitance, and $C_{ch}$ can be calculated as

$$C_{ch} = \frac{\partial Q_{ch}}{W \partial V_g}. \tag{4}$$

$Q_{ch}$ is the central region's total charge, and W is the channel width [41]. The calculated values of $C_g$ for the n- and p-type sub-5-nm 7-9-7-AGNR FET are shown in Tables SI and SII in the Supplemental Material, respectively. It is found that $C_g$ (0.143–0.315 fF/μm for n-type device, 0.127–0.247 fF/μm for p-type device) of the 7-9-7-AGNR FET is much smaller than either the HP (0.6 fF/μm) or LP (0.69 fF/μm) ITRS standard. Figures 7(a) and 7(b) additionally show the values of $\tau$ of n-type and p-type sub-5-nm

7-9-7-AGNR FETs as a function of gate length. It is found that the intrinsic delay time of both the n-type and p-type FETs with various $L_g$ and $L_{UL}$ can fulfill the ITRS requirement (0.423 ps) for the HP devices. Most of the intrinsic delay time can also meet the ITRS LP standard (1.493 ps) for the LP devices (except for the one in the n-type FET with $L_g$ = 3 nm and $L_{UL}$ = 5 nm). Moreover, the smallest value of $\tau$ under certain $L_g$ and $L_{UL}$ reaches 10 times smaller than the HP/LP ITRS standards, showing the great potential applications of 7-9-7-AGNR in high-switching-speed FET devices.

Another significant concern on FET applications is the switching-energy cost by PDP, which can be calculated by

$$PDP = V_{dd}I_{on}\tau = C_g V_{dd}^2. \tag{5}$$

As shown in Figs. 7(c) and 7(d), in the p-type 7-9-7-AGNR FETs with a fixed $L_g$, PDP monotonously decreases with increasing $L_{UL}$. Whereas, the smallest PDP is likely to appear with $L_{UL}$=1 nm for various $L_g$ values in the n-type FETs. Note that the PDPs of both n-type (0.018–0.008 fJ/μm) and p-type (0.013–0.009 fJ/μm) sub-5-nm FETs are much lower than the ITRS requirements for HP (0.24 fJ/μm) and LP (0.28 fJ/μm) standards. This feature shows that the 7-9-7-AGNR FET devices hold the advantage of low-power consumption.

### E. Discussion

Before conclusion, we compare of the main parameters, namely, on-state current, subthreshold swing, delay time, and power-delay product of 7-9-7-AGNR FETs with the various low-dimensional FETs of $L_g \leq$ 5 nm. As shown in Table I, the n-type HP 7-9-7-AGNR FET has the largest $I_{on}$ (2423 μA/μm) among the low-dimensional n-type devices reported so far. As for the p-type HP applications, it also has the second largest $I_{on}$ (4277 µA/µm) that is only smaller than 2D phosphorene (4500 µA/µm). In the LP case shown in Table II, the 7-9-7-AGNR FET can hold remarkably large $I_{on}$, i.e., 1988 µA/µm and 920 µA/µm that correspond to the No. 2 (inferior to β-TeO$_2$) and No. 3 (inferior to β-TeO$_2$ and WSe$_2$) largest current for the n-type and p-type devices, respectively. Nonetheless, the 7-9-7-AGNR FET can have obviously smaller SS than that of phosphorene, β-TeO$_2$ and WSe$_2$, showing a better gate-control ability. In addition, the 7-9-7-AGNR FET presents the lowest $\tau$ and PDP, which further show its promising potential for the LP and HP device applications.

TABLE I. Comparison of the upper-performance limit of the one-dimensional 7-9-7-AGNR FET with other MOSFETs of Lg ≤ 5 nm for HP applications.

|  | Doping type | $I_{ON}$(µA/µm) | SS(mV/dec) | $\tau$ (ps) | PDP(fJ/µm) |
|---|---|---|---|---|---|
| MoS$_2$[44] | n-type | 473 | 58 | 1.287 | 0.195 |
|  | p-type | 440 | 46 | 0.396 | 0.096 |
| WSe$_2$[15] | p-type | 1464 | 82 | 0.168 | 0.156 |
| Bi$_2$O$_2$Se[16] | n-type | 916 | 114 | 0.240 | 0.141 |
|  | p-type | 585 | 96 | 0.375 | 0.140 |
| BiH[15] | p-type | 2320 | 77 | 0.020 | 0.029 |
| GeSe[41] | n-type | 518 | 130 | 0.124 | 0.041 |
|  | p-type | 1703 | 60 | 0.054 | 0.059 |
| Phosphorene[23] | p-type | 4500 | 76 | 0.055 | 0.135 |
| Tellurene[13] | p-type | 2114 | 102 | 0.068 | 0.098 |
| Silicane[10] | n-type | 1374 | 65 | 0.042 | 0.037 |
|  | p-type | 871 | 67 | 0.075 | 0.043 |
| Arsenene[8] | p-type | 2030 | 77 | 0.017 | 0.032 |
| MoSi$_2$N$_4$[17] | n-type | 1390 | 44 | 0.064 | 0.057 |
|  | p-type | 618 | 64 | 0.140 | 0.055 |
| β-TeO$_2$[45] | n-type | 670 | 84 | 0.034 | 0.064 |
|  | p-type | 3700 | 96 | 0.074 | 0.090 |
| 7-9-7-AGNR | n-type | 2423 | 65 | 0.006 | 0.008 |
|  | p-type | 4277 | 54 | 0.004 | 0.010 |

TABLE II. Comparison of the upper-performance limit of the one-dimensional 7-9-7-AGNR FET with other MOSFETs of $L_g \leq 5$ nm for LP applications.

| | Doping type | $I_{ON}$ (μA/μm) | SS (mV/dec) | τ (ps) | PDP (fJ/μm) |
|---|---|---|---|---|---|
| $MoS_2$[44] | n-type | 324 | 56 | 0.552 | 0.093 |
| | p-type | 425 | 46 | 0.411 | 0.075 |
| $WSe_2$[15] | p-type | 1132 | 63 | 0.149 | 0.108 |
| $ReS_2$[12] | p-type | 329 | 72 | 0.700 | 0.150 |
| BiH[15] | p-type | 179 | 67 | 0.168 | 0.018 |
| GeSe[41] | n-type | 274 | 90 | 0.320 | 0.055 |
| Phosphorene[23] | p-type | 857 | 85 | 0.193 | 0.108 |
| Tellurene[13] | p-type | 451 | 57 | 0.206 | 0.063 |
| Silicane[10] | n-type | 467 | 77 | 0.054 | 0.016 |
| | p-type | 378 | 67 | 0.136 | 0.012 |
| Arsenene[8] | p-type | 341 | 77 | 0.101 | 0.023 |
| $MoSi_2N_4$[17] | n-type | 1025 | 44 | 0.086 | 0.057 |
| | p-type | 355 | 70 | 0.265 | 0.060 |
| $β-TeO_2$[45] | n-type | 2450 | 78 | 0.025 | 0.076 |
| | p-type | 3600 | 75 | 0.034 | 0.011 |
| 7-9-7-AGNR | n-type | 1988 | 65 | 0.010 | 0.009 |
| | p-type | 920 | 54 | 0.020 | 0.010 |

## IV. CONCLUSION

Based on the he ab initio quantum-transport calculations, we show that the recently synthesized 7-9-7-AGNR is an ideal candidate for the high-performance 1D FETs. We find that the marginal benzene rings in 7-9-7-AGNR play an important role in regulating the transport channel, making 7-9-7-AGNR hold remarkable on-state current. The further calculations on 7-9-7-AGNR FET devices show that the on-state current can be effectively manipulated by the either the gate length or the underlap length. The optimized $I_{on}$ in n-type (p-type) device can reach up to 2423 (4277) and 1988 (920) μA/μm for high-performance and low-power applications, respectively. The large $I_{on}$ values are in the first class among the low-dimensional FETs, which can well satisfy the ITRS requirements. Moreover, the 7-9-7-AGNR FET has excellent gate-control ability with quite small SS, and the lowest delay time and power-delay product among the low-dimensional FETs. These findings reveal the great potential applications of the 7-9-7-AGNR based FETs in the high-speed and low-power consumption chips.

**Acknowledgments**

This work was supported by the National Natural Science Foundation of China (No. 12074301).

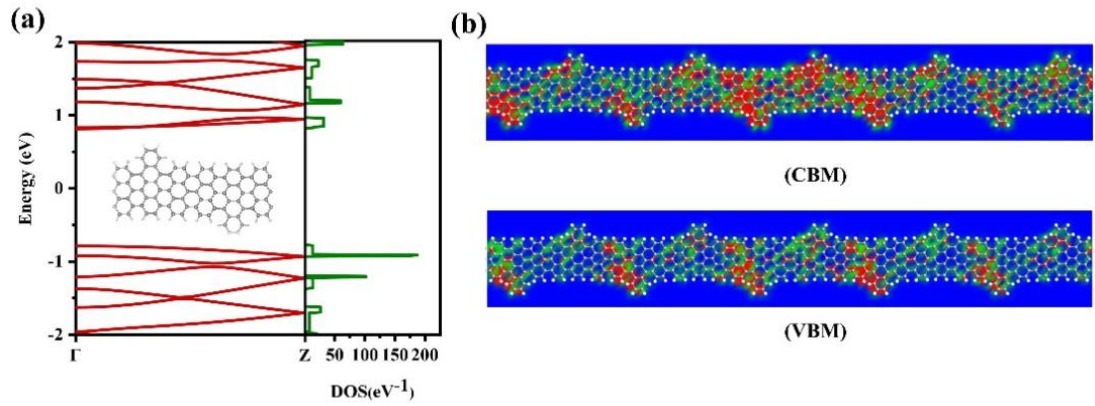

Fig.1 (a) Calculated electronic band structure (left panel) of 7-9-7-AGNR and density of states (right panel). The inset in the left panel shows the atomic structure of 7-9-7-AGNR within a unit cell. (b) Transmission eigenstates distribution of 7-9-7-AGNR with $V_{bis}$ = 0.64V and $V_g$ = 0V.

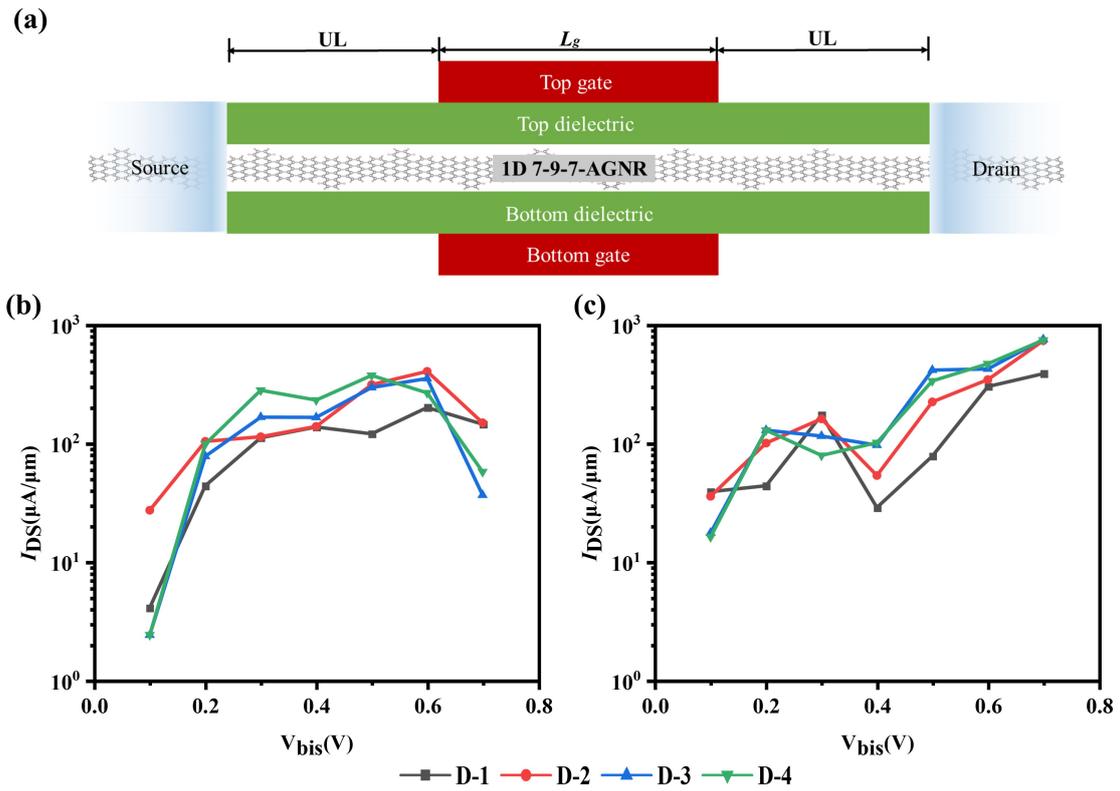

Fig.2 (a) Schematic diagram of the double-gated (DG) one-dimensional 7-9-7-AGNR FET. (b) and (c) show I-$V_{bis}$ characteristics of n- and p-type FETs, respectively, with different $V_{bis}$ from 0.0 V to 0.7 V (D-1,D-2,D-3, and D-4 represent the doping concentration of $1\times10^{13}$, $2\times10^{13}$, $3\times10^{13}$, $4\times10^{13}$ cm$^{-2}$, respectively).

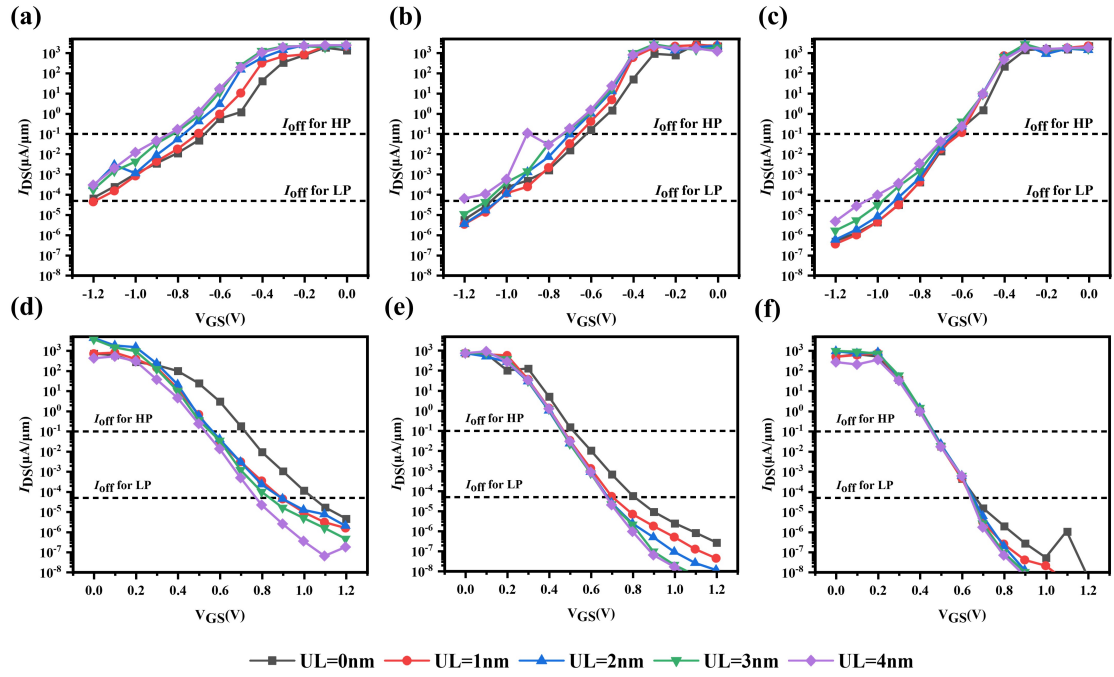

Fig.3 $I_{DS}$-$V_g$ characteristics for the n-type and p-type 7-9-7-AGNR FETs with $V_{bis}$ = 0.6 V. (a-c) are the n-type FETs with different gate lengths (1nm, 3nm, 5nm) and $L_{UL}$. (d-f) present the p-type FETs with different gate lengths (1nm, 3nm, 5nm) and $L_{UL}$.

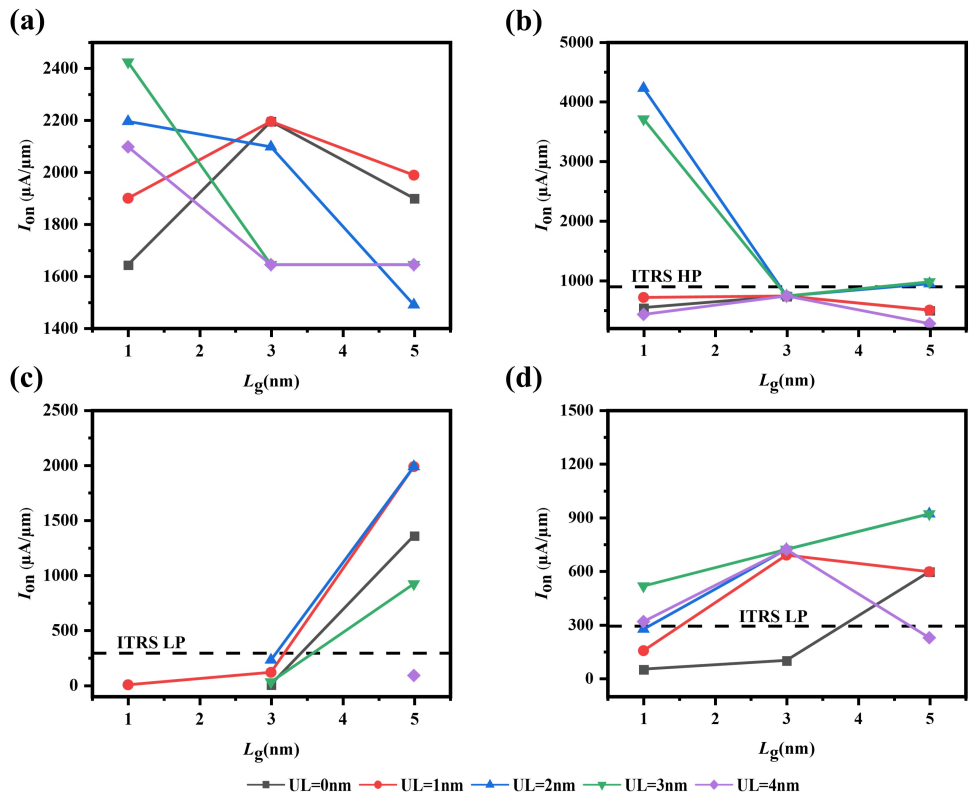

Fig.4 (a)–(d) On-state current as a function of gate length for n-type (a, c) and p-type (b, d) 7-9-7-AGNR FETs with different $L_{UL}$ for the HP (a, b) and LP (c, d) applications, respectively. Black dashed lines represent the ITRS HP and LP requirements.

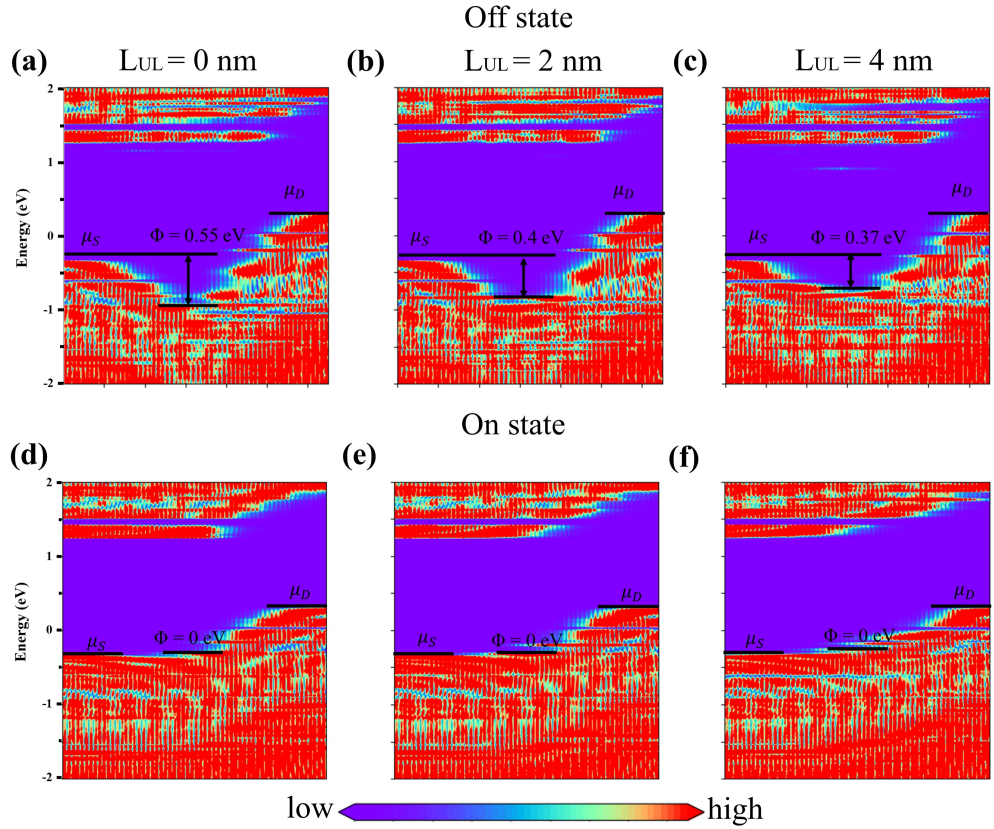

Fig.5 Spatially resolved LDOS for p-type 7-9-7-AGNR FETs in the off (a)–(d) and on states (e)–(h) with 1-nm gate length. (a), (b), (c) represent the LDOS of off state ($I_{on} = 5 \times 10^{-5}$ μA/μm) with underlap lengths of 0 nm, 2 nm, and 4 nm, respectively. (d), (e), (f) correspond to the on-state ($I_{on}$ = 50.44, 273.97, 316.6 μA/μm) LDOS with underlap lengths of 0 nm, 2 nm, and 4 nm, respectively. $μ_S$ and $μ_D$ are the electrochemical potentials of the source and drain, respectively.

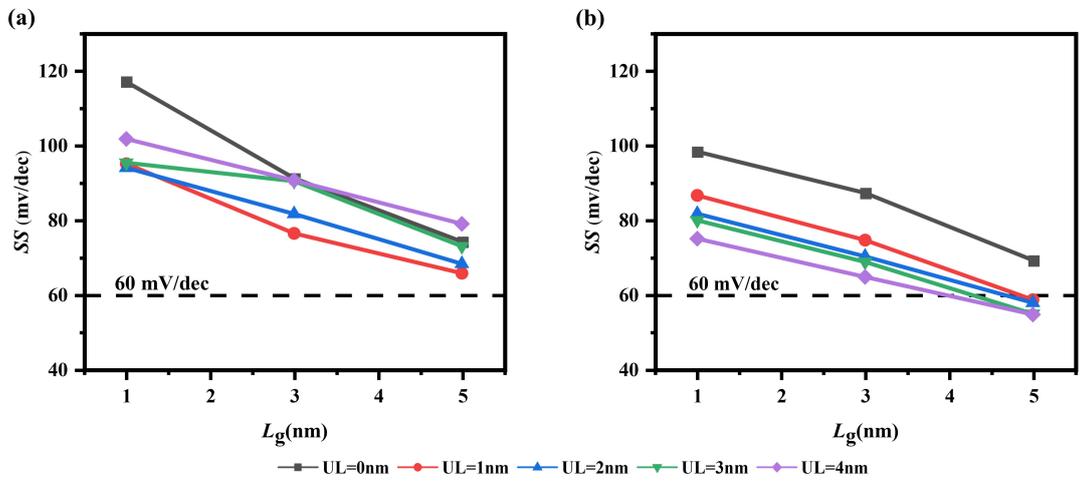

Fig.6 SS as a function of gate length for (a) n-type and (b) p-type 7-9-7-AGNR FETs with different $L_{UL}$. Black dashed lines indicate the Boltzmann limit of 60 mV/dec for SS at room temperature.

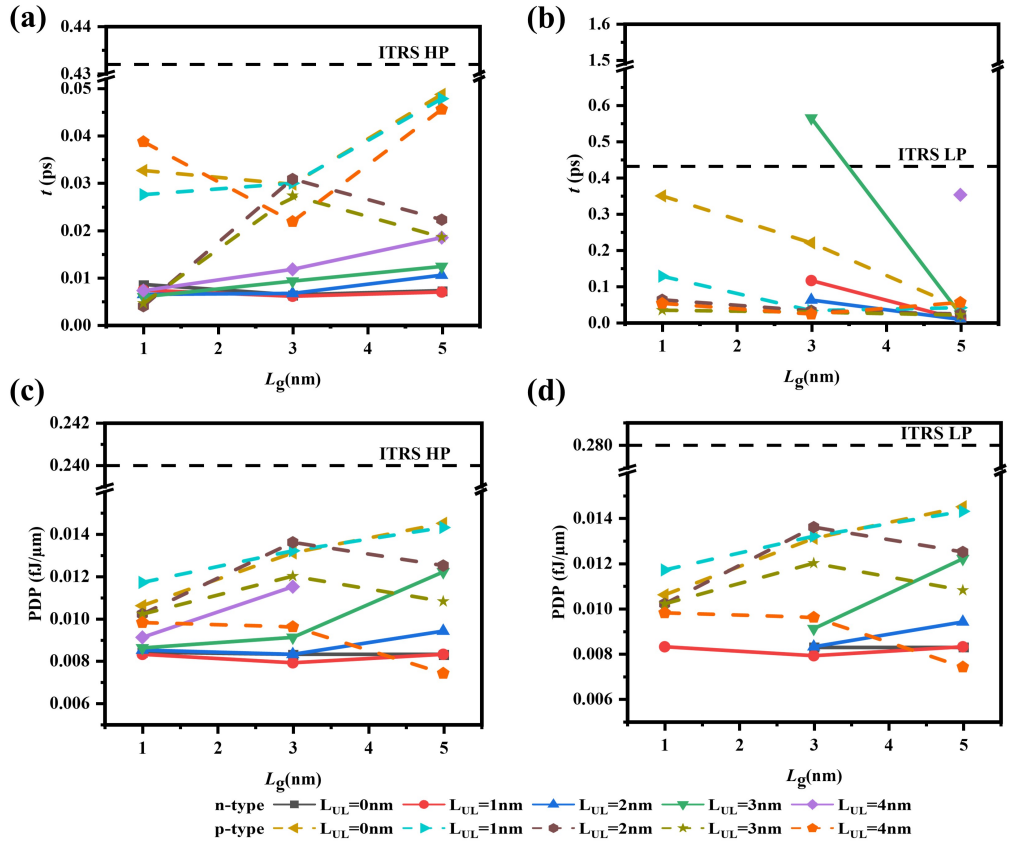

Fig.7 Intrinsic delay time (τ) (a),(b) and PDP (c),(d) as a function of gate length for n-type and p-type FETs with different $L_g$. Black dashed lines are the ITRS HP and LP requirements for τ and PDP, respectively.